\documentclass[journal,article,submit,moreauthors,pdftex,fluids]{Definitions/mdpi} 

\usepackage{graphicx}
\usepackage{amsmath,amssymb}\usepackage{bm}
\usepackage{ascmac}
\usepackage{theorem}
\usepackage{mathtools}
 \usepackage{txfonts}
\mathtoolsset{showonlyrefs}
\mathtoolsset{centercolon}

\newcommand{\R}{\varmathbb{R}}

\newcommand{\inta}{\int_{\R^3} \int_0^{+\infty}}
\newcommand{\dA}{\, \varphi(\mathcal{I}) \, d\mathcal{I} d {\boldsymbol C}}
\newcommand{\da}{\, \varphi(\mathcal{I}) \, d\mathcal{I} d {\boldsymbol \xi}}
\newcommand{\cc}{{\boldsymbol \xi}}
\newcommand{\xx}{\mathbf{x}}

\newcommand{\PP}{\mathcal{P}}

\newcommand{\difg}[2]{\ifx#2\rho{#1_{\rho}}\else{#1_{\hat{G}_{ll}}}\fi}
\newcommand{\dift}[2]{\ifx#2\rho{\left(\frac{\partial #1}{\partial \rho}\right)_T}\else{\left(\frac{\partial #1}{\partial T}\right)_\rho}\fi}

\firstpage{1} 
\pubvolume{xx}
\issuenum{1}
\articlenumber{5}
\pubyear{2020}
\copyrightyear{2020}
\history{Received: date; Accepted: date; Published: date}

\Title{Molecular Extended Thermodynamics of Rarefied Polyatomic Gases with a new  Hierarchy of Moments}

\Author{Takashi Arima $^{1,\ddagger}$\orcidA{} and Tommaso Ruggeri $^{2,\ddagger}$*\orcidB{} }

\AuthorNames{Takashi Arima and Tommaso Ruggeri}

\address{%
$^{1}$ \quad Department of Engineering for Innovation National Institute of Technology, Tomakomai College, Tomakomai, Japan; arima@tomakomai-ct.ac.jp\\
$^{2}$ \quad Department of Mathematics and Alma Mater Research Center on Applied Mathematics AM$\,^2$, University of Bologna, Bologna, Italy; tommaso.ruggeri@unibo.it}

\corres{Correspondence: tommaso.ruggeri@unibo.it}

\secondnote{These authors contributed equally to this work.}

\abstract{
Recently, Pennisi and Ruggeri [J Stat Phys 179, 231–246 (2020)] consider the classical limit of the relativistic theory of moments associated with the  Boltzmann-Chernikov equation truncated at a tensorial index $N+1$ and they proved that there exists a unique possible choice of the moments in the classical case for a given $N$ both for monatomic and polyatomic gases. In particular, in polyatomic gases, there exists a new hierarchy of moments that is more general than the one considered in the recent literature. As consequence, when $N=2$, in the classical limit, there is a theory with $15$ fields. In this paper, we consider this system of moments, and we close the system using the maximum entropy principle. It is shown that the theory contains as a principal subsystem the previously polyatomic $14$ fields theory, and in the monatomic limit, in which the dynamical pressure vanishes,  the differential system converges instead to  Grad  13-moments system to the   14 moments theory proposed by Kremer [Annales de l'I.H.P. Physique théorique,  45, 419-440 (1986)]. 
}


\begin{document}

\section{Introduction}

 It is well known that when the Knudsen number $K_n$  is very high,  the appropriate  theory of the monatomic gas is the Boltzmann equation\footnote{As usual, the repeated indices denote the summation.} 
 \begin{equation} \label{eq:Boltzmann}
 	\frac{\partial f}{\partial t} + \xi_i\, \frac{\partial  f}{\partial x_i} = Q(f),
 \end{equation}
 where the state of the gas can be described by the distribution function $f(\xx,\cc, t)$, 
being $\xx \equiv(x_i),\cc\equiv(\xi_i),t$ the space coordinates, the microscopic velocity and the time, respectively, and $Q$ denotes the collisional term.
A huge literature exists on the Boltzmann equation 
  in which very important mathematical contributions were given by Cercignani \cite{Cer1,Cer2}.
   Associating to the distribution function, we can construct macroscopic observable quantities that are called moments ($m $ is the atomic mass)\footnote{When $n=0$, we have the mass density $F=\rho$. }:
 \begin{equation*}
 F_{k_1 k_2 \dots k_n}(\xx,t) = m \int_{\R^3} f(\xx,\,\bm{\xi},t)  \, \xi_{k_1}  \xi_{k_2} \dots  \xi_{k_n} \, d\cc , \qquad k_1,k_2, \dots, k_n= 1,2,3 \quad \text{and} \quad n=0,1,2,\dots
 \end{equation*}
As a consequence of the  Boltzmann equation \eqref{eq:Boltzmann}, we have   an infinite hierarchy of moment equations that are in the form of balance laws:
\begin{equation}\label{momentini}
\frac{\partial F_{k_1 k_2 \dots k_n}}{\partial t}+ \frac{\partial F_{k_1 k_2 \dots k_n k_{n+1}}}{\partial x_{k_{n+1}}} = P_{k_1k_2 \dots k_n}, \qquad n=0,1,\dots
\end{equation} 
where
\begin{equation*}
P_{k_1k_2 \dots k_n} = m \int_{\R^3} Q(f)\, \xi_{k_1}  \xi_{k_2} \dots  \xi_{k_n}  \, d\cc, \qquad P=P_{k_1}=P_{kk}=0.
\end{equation*}

  Instead, for small $K_n$, the continuum approach with the classical constitutive equations of Navier--Stokes and Fourier (NSF)  gives a satisfactory theory and is applicable for a more  large class of fluids, such as polyatomic  and dense gases.

Beyond the assumption of the local thermodynamic equilibrium which determines the application range of the the NSF theory, the Rational Extended Thermodynamics (RET)  has been developed  \cite{RET,book,newbook}.  In the theory, dissipative fluxes, such as viscous stress and heat flux, are adopted as independent variables in addition to the usual hydrodynamic variables, and we assume a system of balance equations
 with local-type constitutive equations.
More precisely, the main idea of RET is to consider for sufficient large  Knudsen number a structure of balance laws that have the form dictated by the moments \eqref{momentini} truncated at some level. The main problem is in this case that we need a closure procedure. The first approach was pure phenomenological which adopts the structure of moments but forgets that the $F's$ are moments of a distribution function. The phenomenological closure was  obtained by using the universal principles of continuum thermomechanics---(I) \emph{the Galilean invariance and the objectivity principle}, (II) \emph{the entropy principle}, and (III) \emph{the causality and thermodynamic stability (i.e., convexity of the entropy)}---to select admissible constitutive equations.

 The first paper with this procedure  was given by Liu and M\"uller \cite{LiuMul}  motivated by a paper of Ruggeri \cite{Acta} considering $13$ moments of the form \eqref{momentini} with $n=0,1,2,3$ and taking only  the trace of the triple tensor with respect two indexes: $\left({F,F_{k_1},F_{k_1 k_2},F_{k_1 k k}}\right)$.
  It was surprising that the macroscopic  closure  obtained only by adopting the previous universal principles gives the same    system obtained by  Grad   \cite{Grad} using a complete different  closure at kinetic level.  Successively, Kremer presented a refined model  
 with $14$ fields (monatomic ET$_{14}$)  \cite{Kremer14} by adopting a new scalar field $F_{jjkk}$ in addition to the previous $13$ fields: $\left({F,F_{k_1},F_{k_1 k_2},F_{k_1 k k},F_{jjkk}}\right)$.

 For the case with many fields such as \eqref{momentini} truncated at a tensorial order $\bar{N}$, to avoid the complexity of the phenomenological approach, the so-called \emph{molecular extended thermodynamics} has been proposed  in which the macroscopic quantities are moments of the distribution function \cite{ET}. For the closure, we adopt as technique the variational procedure of  Maximum Entropy Principle (MEP)  introduced first in the theory of moments by  Kogan \cite{Kogan} and resumed in $13$ moments by Dreyer \cite{Dreyer} and for many  moments  in the first edition of M\"uller and Ruggeri book  in which it was proved that the closed system is symmetric hyperbolic \cite{ET}. See also on this subject the contribution of Boillat and Ruggeri   \cite{Boillat-1997} in which they proved that in the molecular approach the MEP closure is equivalent to the closure using an entropy principle.
 
 The first relativistic version of the modern RET was give by Liu, M\"uller and Ruggeri (LMR) \cite{LMR} considering the Boltzmann-Chernikov relativistic equation
 \cite{BGK,Synge,KC}:
 \begin{equation}\label{BoltzR}
 p^\alpha \partial_\alpha f = Q.
 \end{equation}
in which now  the distribution function  $f$ depends      on  $(x^\alpha,  p^\beta)$, where $x^\alpha$ are the space-time coordinates, $p^\alpha$ is the four-momentum $p_\alpha p^\alpha = m^2 c^2$, $\partial_{\alpha} = \partial/\partial x^\alpha$, $c$ denotes the light velocity, $m$ the mass in the rest frame and $\alpha, \beta =0,1,2,3$.
 	The relativistic moment equations associated with \eqref{BoltzR}, truncated at tensorial index $N+1$, are now\footnote{When $n=0$, the tensor reduces to $A^\alpha$. Moreover, the production tensor in the right-side of \eqref{RelmomentMono} is zero for $n=0,1$,  because the first $5$ equations   represent the conservation laws of the particles number and the energy-momentum, respectively.}:
 \begin{equation}\label{Relmomentseq}
 \partial_\alpha A^{\alpha \alpha_1 \cdots \alpha_n  } =  I^{  \alpha_1 \cdots \alpha_n   }
 \quad \mbox{with} \quad n=0 \, , \,\cdots \, , \,  N 
 \end{equation}
 with  
 \begin{align}\label{RelmomentMono}
 A^{\alpha \alpha_1 \cdots \alpha_n  } = \frac{c}{m^{n-1}} \int_{\mathbb{R}^{3}}
 f  \,  p^\alpha p^{\alpha_1} \cdots p^{\alpha_n}  \, \, d \boldsymbol{P}, \qquad I^{\alpha_1 \cdots \alpha_n  } = \frac{c}{m^{n-1}} \int_{\mathbb{R}^{3}}
 Q  \,   p^{\alpha_1} \cdots p^{\alpha_n}  \, \, d \boldsymbol{P} ,
 \end{align}
and 
 \begin{equation*}
 d \boldsymbol{P} =  \frac{dp^1 \, dp^2 \,
 	dp^3}{p^0} .
 \end{equation*}
 When $N=1$, we have the relativistic Euler system, and when $N=2$, we have the LMR theory of a relativistic gas with $14$ fields \footnote{In the monatomic case, from \eqref{RelmomentMono}, we have $A^{\alpha \beta}_{\,\,\, \,\beta} = c^2 A^\alpha$ and therefore only $14$ equations of \eqref{Annals} are independent.}:
 \begin{equation}
  \partial_\alpha A^{\alpha } = 0, \quad \partial_\alpha A^{\alpha \beta} =0, \quad  \partial_\alpha A^{\alpha \beta \gamma} =  I^{  \beta \gamma  }, \qquad \left(\beta,\gamma=0,1,2,3; \,\, I^\alpha_{\,\,\alpha} =0 \right). \label{Annals}
  \end{equation}
 The surprising results was  that the LMR theory     converges, in the classical limit, to the monatomic ET$_{14}$ theory by Kremer for monatomic gas not the Grad theory (ET$_{13}$) as was expected \cite{Weiss-Dreyer,RET,PRS}. 
 
 For many years, the applicability range of RET was only limited to monatomic gases both in the classical and relativistic regime. For rarefied polyatomic gases, after some previous tentatives \cite{Liu,KremerPoly},  Arima, Taniguchi, Ruggeri and Sugiyama \cite{Arima-2011} proposed a binary hierarchy of field equations with $14$ fields (polyatomic ET$_{14}$) because now there is also, as a new field, the dynamical pressure 
  relating to the relaxation of the molecular internal modes which is identically to zero in monatomic gases:
  \begin{align} \label{ET14poly}
    & \frac{\partial F}{\partial t} +  \frac{\partial F_i}{\partial x_i} =0, \nonumber\\
    & \frac{\partial F_j}{\partial t} +  \frac{\partial F_{ij}}{\partial x_i} =0,\nonumber\\
    & \frac{\partial F_{ij}}{\partial t} +  \frac{\partial F_{ijk}}{\partial x_k} =P_{ ij},
    && \frac{\partial G_{ll}}{\partial t} +  \frac{\partial G_{llk}}{\partial x_k} = 0, \\
    & && \frac{\partial G_{lli}}{\partial t} +  \frac{\partial G_{llik}}{\partial x_k} =Q_{lli}.\nonumber
  \end{align}
  where $F(=\rho)$ is the mass density, $F_i(=\rho v_i)$ is the momentum density, $G_{ll}=\rho v^2 +2 \rho\varepsilon$ is two times  the energy density, $F_{ij}$ is the momentum flux, and
  $G_{llk}$ is the energy flux. As usual $v_i$ denotes the components of velocity and $\varepsilon $ is the internal energy.
   $F_{ijk}$ and $G_{llik}$ are the fluxes of $F_{ij}$ and $G_{lli}$,  respectively, and $P_{ij}$ ($P_{ll}\neq 0$) and $Q_{lli}$  are the productions with respect to $F_{ij}$ and  $G_{lli}$, respectively.
  In the parabolic limit, this theory converges to the NSF theory, and in the monatomic  singular limit, it converges to the Grad system \cite{Arima-2013,book,newbook}.
  This hierarchy was justified at kinetic level in \cite{Pavic-2013,Ruggeri-2020RdM,Arima-2014}
   using  the same form of Boltzmann equation \eqref{eq:Boltzmann} but with a distribution function $f\left(\xx,\cc, t, \mathcal{I}\right)$ that depends on a non-negative \textit{internal energy parameter} $\mathcal{I}$, that takes into account the influence of the internal degrees of freedom of a molecule on energy transfer during collisions \cite{Borgnakke-1975,Bourgat-1994}.  
The theory with many moments was also developed in \cite{Pavic-2013,Arima-2014,Arima-2016}:
\begin{align}\label{momentinipoli}
 \begin{split}
 \frac{\partial F_{k_1 k_2 \dots k_n}}{\partial t}+ \frac{\partial F_{k_1 k_2 \dots k_n k_{n+1}}}{\partial x_{k_{n+1}}} = P_{k_1k_2 \dots k_n}, &\\ 
&\quad \frac{\partial G_{llk_1 k_2 \dots k_m}}{\partial t}+ \frac{\partial G_{llk_1 k_2 \dots k_n k_{m+1}}}{\partial x_{k_{m+1}}} = Q_{llk_1k_2 \dots k_m}, 
 \end{split}
\end{align}
with the following definition of moments of polyatomic gases ($\xi^2=|\bm{\xi}|^2 =\xi_j\xi_j$):
\begin{align*}
\begin{split}
& F_{k_1k_2\dots k_n}=m \inta f(\xx,\cc,t,\mathcal{I})\, \xi_{k_1}\xi_{k_2} \dots \xi_{k_n} \, \varphi(\mathcal{I}) \, d\mathcal{I} \, d\cc, \\
& P_{k_1k_2\dots k_n}=m \inta Q(f) \xi_{k_1}\, \xi_{k_2} \dots \xi_{k_n} \, \varphi(\mathcal{I}) \, d\mathcal{I}  \, d\cc, \\
&	G_{llk_1k_2\dots k_m}  = \inta \left(m \xi^2 + 2\mathcal{I} \right)f(\xx,\cc,t,\mathcal{I})\, \xi_{k_1}\xi_{k_2} \dots \xi_{k_m} \, \varphi(\mathcal{I}) \, d\mathcal{I}  \, d\cc,\\
&	Q_{llk_1k_2\dots k_m}  = \inta \left(m \xi^2 + 2\mathcal{I} \right)Q(f) \, \xi_{k_1}\xi_{k_2} \dots \xi_{k_m} \, \varphi(\mathcal{I}) \, d\mathcal{I}  \, d\cc ,
\end{split}
\end{align*}
where $\varphi(\mathcal{I})$ is the state density corresponding to $\mathcal{I}$, i.e., $\varphi (\mathcal{I})d\mathcal{I}$ represents the number of internal state between $\mathcal{I}$ and $\mathcal{I} + d \mathcal{I}$. 
 We need to remark that the two blocks of hierarchies in \eqref{momentinipoli}  are not separated because the last fluxes in both hierarchies together with the productions terms are functions to be determined by the closure of all densities $(F_{k_1 k_2 \dots k_n}, 	G_{llk_1k_2\dots k_m})$. The index $n=0,1,\dots,\cal{N}$ and $m=0,1,\dots,\cal{M}$. Moreover, $P=P_{k_1}=Q_{ll}=0$ since the first $4$ equations of the $F$' hierarchy and the first scalar equation of $G$'s hierarchy represent the mass, momentum and energy conservation, respectively. It was studied in \cite{Arima-2014} that the physically meaning full choice of the truncated order $\cal{N}$ and $\cal{M}$, in the sense of the Galilean invariance and the characteristic velocity, is ${\cal{M}}={\cal{N}} - 1$.

Recently, Pennisi and Ruggeri first constructed a relativistic version of polyatomic gas in the case of $N=2$ \cite{Annals}. Then, in \cite{PRS}, they studied the classical limit of generic moments equations \eqref{Relmomentseq} 
for a fixed $N$ both in monatomic gas of which moments are \eqref{RelmomentMono} and in polyatomic gas of which moments are given by:
\begin{align*} 
\begin{split}
& A^{\alpha \alpha_1 \cdots \alpha_n  } = \frac{c}{m^{n-1}} \int_{\mathbb{R}^{3}}
\int_0^{+\infty} f  \,  p^\alpha p^{\alpha_1} \cdots p^{\alpha_n}  \, \left( 1 +  n \frac{\mathcal{I}}{m \, c^2} \right)\, 
\varphi(\mathcal{I}) \, d \mathcal{I} \, d \boldsymbol{P}  \, , \\
& I^{\alpha_1 \cdots \alpha_n  } = \frac{c}{m^{n-1}} \int_{\mathbb{R}^{3}}
\int_0^{+\infty} Q  \,  p^{\alpha_1} \cdots p^{\alpha_n}  \, \left( 1 +  n \frac{\mathcal{I}}{m \, c^2} \right)\, 
\varphi(\mathcal{I}) \, d \mathcal{I} \, d \boldsymbol{P}, \\
\end{split}
\end{align*}
with a  distribution functions $f(x^\alpha, p^\beta,\mathcal{I})$ depends on the extra energy variable $\mathcal{I}$ similar to the classical one.
They proved that there is a unique possible choice of classical moments for a prescribed truncation index $N$ of \eqref{Relmomentseq}. In particular, for $N=2$, in the monatomic case, we have in the classical limit  the monatomic $14$-moment equations by Kremer \cite{Kremer14} according with the old results of \cite{Weiss-Dreyer}. Instead, in  the polyatomic case, for $N=2$, we have, as classical limit, $15$ moments in which, in addition to the previous polyatomic $14$-moment equations  \eqref{ET14poly}, one equation for a mixed type moment $H_{llmm}$  defined by
\begin{align}\label{hll}
 H_{llmm} = 2G_{llmm}-F_{llmm}
\end{align}
is involved. For many moments,  the new hierarchy contains, in addition to the $F's$ and $G's$ hierarchies \eqref{momentinipoli} with $n=0,1,\dots N$ and $m=0,1,\dots N-1$, more complex $N+1$ hierarchies for  mixed type of moments (see \cite{PRS}). 
For more details on RET beyond the monatomic gas, see the new book  of Ruggeri and Sugiyama \cite{newbook}.

The aim of the present paper is to study the closure of  the most simple case of this new hierarchy, that is the system with $15$ equations (ET$_{15}$):
\begin{align}
\begin{split}
 &\frac{\partial F}{\partial t}+\frac{\partial F_k}{\partial x_k}=0,\\
 &\frac{\partial F_i}{\partial t}+\frac{\partial F_{ik}}{\partial x_k}=0, \\
 &\frac{\partial F_{ij}}{\partial t}+\frac{\partial F_{ijk}}{\partial x_k}=P_{ij}, \ \ \ \ \ \ \ \ \ \ \frac{\partial G_{ll}}{\partial t}+\frac{\partial G_{llk}}{\partial x_k}=0,\\
 &\qquad \qquad \qquad \qquad \qquad \quad   \frac{\partial G_{lli}}{\partial t}+\frac{\partial G_{llik}}{\partial x_k}=Q_{lli},\\
  &\hspace{7.5cm}\frac{\partial H_{llmm}}{\partial t}+\frac{\partial H_{llmmk}}{\partial x_k}=R_{llmm},
\end{split}\label{balanceFG1}
\end{align}
where  $H_{llmmk}$ is the flux of $H_{llmm}$ given by \eqref{hll}, and  $R_{llmm}$ is the production with respect to  $H_{llmm}$.  
In the following, after presenting a equilibrium properties of the distribution function, we close the system \eqref{balanceFG1} by means of MEP. As the collisional term, we introduce the generalized BGK model for a relaxation processes of molecular internal modes \cite{Ruggeri-2020RdM}. We show that the derived closed set of the moment equations involves the polyatomic ET$_{14}$ theory as a principal subsystem, the monatomic ET$_{14}$ theory in the monatomic singular limit, and the NSF theory as its parabolic limit.

\section{Molecular Extended Thermodynamics with ${15}$-field}
First, we recall the equilibrium distribution function for polyatomic gas that was deduced first in the polytropic case ($p,\varepsilon,\rho,T$ denote as usual the equilibrium pressure, the equilibrium specific internal energy, the mass density and the absolute temperature, while $k_B$ is the Boltzmann constant and the constant $D = 3 + f_i$, where $f_i$ are the degree of freedom; in the monatomic gas $D=3$)
\begin{equation}
p=\frac{k_B}{m}\rho T, \qquad \varepsilon = \frac{D}{2} \frac{k_B}{m}T \label{EOSnonpoly}
\end{equation}
 in \cite{Bourgat-1994,Pavic-2013} and in the present case of non polytropic gas 
 \begin{equation}
 p=p(\rho,T)=\frac{k_B}{m}\rho T, \qquad \varepsilon \equiv \varepsilon_E( T) \label{EOS}
 \end{equation}
 in \cite{RuggeriSpiga,Ruggeri-2020RdM}:
 \begin{align}
f_E=f^{K}_E f^{I}_E \label{fE},
\end{align}
where $f^K_E$ is the Maxwellian distribution function and $f^{I}_E$ is the distribution function of the internal mode:
\begin{align}
&f^{K}_E = \frac{\rho}{m}\left(\frac{m}{2\pi k_B T}\right)^{3/2} \exp \left(- \frac{mC^2}{2 k_B T}\right), \qquad
f^{I}_E = \frac{1}{A(T)} \exp \left(- \frac{\mathcal{I}}{k_B T}\right),
 \label{fKI}
\end{align}
with  $A(T)$ is  the normalization factor  (partition function):
\begin{align}\label{partitions}
 A(T) = \int_0^{+\infty} \varphi(\mathcal{I}) \mathrm{e}^{- {\beta_E \mathcal{I}}}\mathrm{d}\mathcal{I}, \qquad \beta_E = 1/(k_B T),
\end{align}
and we have put  with 
$C_i=\xi_i-v_i$ $(C^2=C_j C_j) $ 
 the peculiar velocity.  
 
The specific internal energy is the moment of $f_E$ as follows:
\begin{align}
&\varepsilon = \varepsilon_E(T) = \varepsilon_E^{K}(T) + \varepsilon_E^I(T) = \frac{1}{2\rho}\inta (mC^2 + 2\mathcal{I})f_E \dA,  \label{CaEqSt}
\end{align}
where $\varepsilon_E^K$ and $\varepsilon_E^I$ are the equilibrium kinetic (translational) and internal specific energies defined by
\begin{align}  \label{int1e}
\begin{split}
	&\varepsilon_E^K(T)  = \frac{1}{2\rho} \inta m C^2 f_E \dA  = \frac{1}{2\rho} \int_{\R^3} mC^2 f^K_E \,  d{\boldsymbol C}= \frac{3}{2}\frac{k_B}{m}T, \\
    &\varepsilon_E^I(T) = \frac{1}{\rho} \inta \mathcal{I} f_E \dA = \frac{1}{m}\int_0^{+\infty} \mathcal{I} f^{I}_E \,  d\mathcal{I} =  \frac{k_B}{m}T^2\frac{\mathrm{d} \log A(T)}{\mathrm{d}T},
\end{split}
\end{align}
where the following relation is taking into account by \eqref{fKI}$_2$ and \eqref{partitions}
\begin{align}
\int_{0}^{\infty} f^{I}_E  \, \varphi(\mathcal{I}) \, d I=1. \label{fint}
\end{align}
The partition function is obtained by integrating \eqref{int1e}$_2$ if the caloric equations of state is given. Then the measure $\varphi(\mathcal{I})$ is determined via the inverse Laplace transformation of \eqref{partitions}. Vice versa, if the partition functions $A(T)$ is  given, for example, by a statistical-mechanical analysis, we obtain the equilibrium energies of internal mode from \eqref{int1e}$_2$  (see for more details \cite{Ruggeri-2020RdM}).  

It was proved in \cite{Ruggeri-2020RdM} that
\begin{align}
 \frac{1}{m^2}\int_0^\infty \mathcal{I}^2 f^{I}_E d\mathcal{I} = \frac{p^2}{\rho^2}\hat{c}_v^I + \varepsilon_E^{I}(T)^2,  \label{cvI}
\end{align}
where 
\begin{align*}
\hat{c}_v^I = m \frac{c_v^I}{k_B}, \quad \text{and} \quad   c_v^I = \frac{d\varepsilon^I_E(T)}{dT}
\end{align*}
is  the specific heat of the internal mode. We remark that the relation between the pressure and the translational internal energy is as follows:
\begin{align*}
p = \frac{2}{3}\rho \varepsilon_E^K(T). 
\end{align*}
The specific entropy density in equilibrium is expressed by
\begin{align*}
s = s(\rho,T) = s^K(\rho, T) + s^I(T),
\end{align*}
with its translational part $s^K$ and internal part $s^I$ which are given by
\begin{align*}
\begin{split}
s^K (\rho,T)&\equiv - \frac{k_B}{\rho} \inta f_E \log f^{K}_E \, \dA, \\
&= \frac{k_B}{m}\log \left(\frac{{T}^{3/2}}{\rho}\right) + \frac{\varepsilon_E^K(T)}{T} - \frac{k_B}{m} \log \left[\frac{1}{m}\left(\frac{m}{2\pi k_B}\right)^{3/2}\right], \\
s^I (T) &\equiv - \frac{k_B}{\rho} \inta f_E \log f^{I}_E \, \dA, \\
&= \frac{k_B}{m}\log A(T) + \frac{\varepsilon_E^I(T)}{T}.
\end{split}
\end{align*}

\subsection{System of balance equations for $15$ fields}

The macroscopic quantities in \eqref{balanceFG1} are defined as the moments of  $f$ as follows:
\begin{align}
 &\left( \begin{array}{l}
  F\\ F_i\\ F_{ij}  \\ F_{ijk}
		\end{array}\right)
 =
 \inta m
  \left( \begin{array}{c}
  1 \\ \xi_i\\ \xi_i \xi_j \\ \xi_i \xi_j \xi_k
		\end{array}\right)
 f \da ,  \nonumber\\
  &\left( \begin{array}{l}
  G_{ll}\\  G_{lli} \\ G_{llik}
		\end{array}\right)
 =
 \inta (m\xi^2 + 2\mathcal{I})
  \left( \begin{array}{c}
  1 \\ \xi_i \\ \xi_i \xi_k
		\end{array}\right)
 f \da, \label{moments} \\ 
  &\left( \begin{array}{l}
  H_{llmm}\\  H_{llmmi}
		\end{array}\right)
  =
 \inta (m\xi^2 + 4\mathcal{I}) \, \xi^2
  \left( \begin{array}{c}
  1 \\ \xi_i 
		\end{array}\right)
 f \da , \nonumber
\end{align}
and the production terms
\begin{align}\label{productioni}
 \mathbf{f}  \equiv &\left( \begin{array}{l}
  P_{ij} \\ Q_{lli} \\ R_{llmm}
		\end{array}\right)
 =
 \inta 
  \left( \begin{array}{c}
  m\xi_i \xi_j \\  (m\xi^2 +2\mathcal{I})\xi_i \\ (m\xi^2 + 4\mathcal{I}) \, \xi^2
		\end{array}\right)
 Q(f) \da .
\end{align}

Since the intrinsic (velocity independent) variables are the moments in terms of the peculiar velocity $C_i$ instead of $\xi_i$, the velocity dependence of the densities is obtained as follows:
\begin{align} \label{density15}
 \begin{split}
  & F=\rho , \\
  &F_i=\rho v_i, \\
  & F_{ ij} = \hat{F}_{ij} + \rho v_{ i}v_{j},  \\
  & G_{ll}= \hat{G}_{ll}+ \rho v^2 , \\
  & G_{lli}  = \hat{G}_{lli} + \hat{G}_{ll}v_i + \hat{F}_{li}v_{l} + \hat{F}_{ll}v_{i} + \rho v^2 v_i,\\
  & H_{llmm} = \hat{H}_{llmm} + 4 \hat{G}_{lli}v_i + 2 \hat{G}_{ll}v^2 + 4\hat{F}_{ij}v_iv_j  + \rho v^4,
 \end{split}
\end{align}
where a hat on a quantity indicates its velocity independent part. 
The conventional fields, i.e.,
\begin{align}
 &\text{mass density}: && \rho = \inta m f \, \dA  = \inta m f_E  \, \dA, \nonumber \\
  &\text{velocity}: && v_i = \frac{1}{\rho}\inta m \xi_i f \, \da = \frac{1}{\rho}\inta m \xi_i  f_E \, \da  , \nonumber\\
  &\text{specific internal energy density}: && \varepsilon =  \varepsilon^K + \varepsilon^I = \frac{1}{2\rho}\inta (mC^2+2\mathcal{I}) f \, \dA ,    \nonumber \\
 &\text{specific translational energy density}: && \varepsilon^K = \frac{1}{2\rho}\inta mC^2 f  \, \dA, \nonumber\\
 &\text{specific internal energy density}: && \varepsilon^I = \frac{1}{\rho}\inta \mathcal{I} f \, \dA , \label{conventional} \\ 
  &\text{total nonequilibrium pressure}: && \PP = \frac{2}{3} \rho \varepsilon^K = \frac{1}{3}\inta m C^2 f \, \dA, \nonumber\\ 
   &\text{dynamic pressure}: && \Pi = \PP - p= \frac{1}{3}\inta m C^2 (f-f_E) \, \dA , \nonumber \\
   &\text{shear stress}: && \sigma_{\langle ij\rangle} = - \inta m C_{\langle i} C_{j\rangle}\,  f \, \dA , \nonumber \\
    &\text{heat flux}: && q_i = \frac{1}{2} \inta \left(m C^2 + 2\mathcal{I} \right)C_i \, f \, \dA , \nonumber
\end{align}
are related to the intrinsic moments as follows:
\begin{align}
 \hat{G}_{ll} = 2\rho \varepsilon  = 2 \rho (\varepsilon^K + \varepsilon^I), \quad \hat{F}_{ll} = 3\PP=3 (p+\Pi), \quad \hat{F}_{\langle ij\rangle} = -\sigma_{\langle ij\rangle}, \quad \hat{G}_{lli}= 2q_i, \label{energies}
\end{align}
where the temperature of the system $T$ is introduced through the caloric equation of state
\begin{align}
	\varepsilon =\varepsilon_E (T). \label{calEOS}
\end{align}
Let us decompose the intrinsic part of $H_{llmm}$ into the equilibrium part and the nonequilibrium part $\Delta$ as follows:
\begin{align*}
	 \hat{H}_{llmm}=  \inta \left(m C^2 + 4\mathcal{I} \right)C^2 f \, \dA = 12 \frac{p^2}{\rho}(5+4y^I) + \Delta,
\end{align*}
where 
\[
y^I = \frac{\rho }{p}\varepsilon^I_E(T)
\]
and  $\Delta$ is defined by
\begin{align}
 \Delta =  \inta \left(m C^2 + 4\mathcal{I} \right)C^2 (f-f_E) \, \dA. \label{defDelta}
\end{align}

Similarly, the velocity dependences of the fluxes and productions are obtained as follows:
\begin{align}
  &F_{ijk} = \hat{F}_{ijk} + \hat{F}_{ij}v_{k} + \hat{F}_{jk}v_{i}+\hat{F}_{ki}v_{j} + \rho v_i v_j v_k ,\nonumber\\
  &G_{llik} = \hat{G}_{llik} + \hat{G}_{lli}v_{k} + \hat{G}_{llk}v_{i}+ 2\hat{F}_{lik}v_l+2\hat{F}_{kl}v_{l}v_{i}+2\hat{F}_{il}v_{l}v_{k} + \hat{F}_{ik}v^2 + \hat{G}_{ll}v_iv_k + \rho v^2 v_i v_k , \nonumber\\
  &H_{llmmk} = \hat{H}_{llmmk} + \hat{H}_{llmm}v_k + 4\hat{G}_{llik}v_i + 2 \hat{G}_{llk}v^2 + 4\hat{G}_{lli}v_iv_k + 4\hat{F}_{ijk}v_i v_j + 2\hat{G}_{ll}v^2 v_k + 4\hat{F}_{ik}v^2 v_i + 4\hat{F}_{ij}v_iv_jv_k + \rho v^4 v_k,\nonumber\\
   &P_{ij}=\hat{P}_{ij}, \label{flussiG} \\
 &Q_{lli}=2v_l \hat{P}_{il}+\hat{Q}_{lli}, \nonumber\\
  &R_{llmm}=4v_iv_j\hat{P}_{ij} + 4v_i \hat{Q}_{lli}+\hat{R}_{llmm}. \nonumber
\end{align}
The velocity dependences  in \eqref{density15} and \eqref{flussiG} take the system \eqref{balanceFG1} Galilean invariant  in agreement with the general theorem on Galilean invariance for a generic   system of balance laws   (see \cite{Ruggeri-1989}). 

The constitutive quantities are now the following moments 
\begin{align*}
   &\hat{F}_{ijk}  = \inta m C_i C_j C_k \, f \, \dA , \\
 &\hat{G}_{llik}  = \inta (mC^2+2\mathcal{I})C_iC_k \,  f \, \dA , \\
 &\hat{H}_{llmmk} = \inta \left(mC^2 + 4\mathcal{I}\right)  C^2 C_k \, f \, \dA,
\end{align*}
that is needed to be determined for the closure of the differential system together with the production terms $P_{ij}, Q_{lli}$ and $ R_{llmm}$.

\subsubsection{Nonequilibrium distribution function derived from MEP}
To close the system \eqref{balanceFG1}, we need the nonequilibrium distribution function $f$, which is derived from the MEP. According with the principle, the most suitable distribution function $f$ of the truncated system \eqref{balanceFG1} is the one that   maximize the entropy 
\begin{align*}
 h =&  -k_B\int_{\R^3} \int_0^{+\infty }f \log f\, \varphi(\mathcal{I})\, d\mathcal{I} d\cc , 
\end{align*}
 under the constraints that the density  moments $F, F_i, F_{ ij},  G_{ll}, G_{lli}, H_{llmm}$ 
  are prescribed as  in \eqref{moments}. Therefore the best approximated distribution function $f_{15}$ is obtained as the solution of a variational problem of the following functional
\begin{align}\label{LL}
 \mathcal{L}\left(f\right)  = &- k_B  \inta f \log f\, \da \nonumber\\
 &+ \lambda \left(F - \inta m  f \da\right)  + \lambda_i \left(F_i - \inta m \xi_i f \da\right)  \\
  & + \lambda_{ij} \left(F_{ij} - \inta m \xi_i \xi_j f \da\right)
 + {\mu} \left( G_{ll} - \inta  \left(m\xi^2 + 2 \mathcal{I}\right) f \da \right)\nonumber\\
 & + \mu_i \left(G_{lli} - \inta  \left(m\xi^2 + 2\mathcal{I} \right) \xi_i  f \da\right)
 + {\zeta} \left( H_{llmm} - \inta \left(m\xi^2 + 4\mathcal{I}\right)\xi^2  f \da \right), \nonumber
\end{align}
where $\lambda$, $\lambda_i$, $\lambda_{ij}$, $\mu$, $\mu_i$, and ${\zeta}$ are the corresponding Lagrange multipliers of the constraints. As  $\mathcal{L}$ is a scalar independent of frame proceeding as in  \cite{Ruggeri-1989}, we can evaluate    the right side of \eqref{LL} in the rest frame of the fluid $(v_i=0)$, and in this way we have  the following   velocity dependence of the Lagrange multipliers (according with the general theorem given in \cite{Ruggeri-1989}):
\begin{align}\label{lamdas}
\begin{split}
 &\lambda = \hat{\lambda} -\hat{\lambda}_iv_i  + \hat{\lambda}_{ij} v_i v_j  + \hat{\mu}v^2 - \hat{\mu}_i v^2 v_i + \hat{\zeta}v^4,\\ 
 &\lambda_i = \hat{\lambda}_i - 2 \hat{\lambda}_{ij}v_j -2 \hat{\mu}v_i + 3\hat{\mu}_i v^2 - 4 \hat{\zeta}v^2v_i,\\
 & \lambda_{ij} = \hat{\lambda}_{ij} - 2\hat{\mu}_i v_i +4\hat{\zeta}v_iv_j,\\
 & \mu = \hat{\mu} - \hat{\mu}_i v_i + 2 \hat{\zeta}v^2,\\ 
 & \mu_i = \hat{\mu}_i - 4 \hat{\zeta}v_i,\\ 
 & \zeta = \hat{\zeta}. 
\end{split}
\end{align}
The~distribution function $f$, which satisfies ${\delta \mathcal{L}}/{\delta f}=0$, is  
\begin{align}\label{fgen}
 \begin{split}
 &f_{15}=\exp\left(-1-\frac{m}{k_B}{\chi}\right), \qquad \text{with}\\
 &{\chi} = \hat{\lambda}+C_i \hat{\lambda}_i + C_iC_j\hat{\lambda}_{ij} + \left(C^2 + \frac{2\mathcal{I}}{m}\right)\hat{\mu}+ \left(C^2+\frac{2\mathcal{I}}{m}\right)C_i\hat{\mu}_i + \left(C^2+\frac{4\mathcal{I}}{m}\right)C^2\hat{\zeta}.
 \end{split}
\end{align}

Taking into account that, in equilibrium, $f_{15}$ coincides with the equilibrium distribution function~\eqref{fE}, we~can easily see that the equilibrium components of the Lagrange multipliers are given by
\begin{align}
{\lambda}_E = \frac{1}{T}\left(- g+ \frac{v^2}{2}\right), \quad {\lambda}_{i_E}= - \frac{v_i}{T}, \quad {\lambda}_{ll_E} = 0, \quad {{\lambda}_{\langle ij\rangle_E}} =0 , \quad \mu_E = \frac{1}{2T}, \quad {{\mu}_{i_E}}=0, \quad  \zeta_E=0, \label{mainE}
\end{align}
where $g (=\varepsilon_E(T) +p/\rho -Ts)$ is the chemical potential. We remark that ${\lambda}_E , {\lambda}_{i_E},  \mu_E$ in \eqref{mainE} are the Lagrange multipliers of the  Euler system, and those are the \emph{main field} symmetrize the Euler system as was proved first by Godunov (see \cite{Godunov,book}).

We observe that the highest power of peculiar velocity in $\chi$ in \eqref{fgen}$_2$  is even, i.e., $C^4$. The highest power is same with the highest tensorial order of the system, and it is revealed in \cite{PRS} that the highest tensorial order of the system obtained in the classical limit is always even, i.e., $2N$. This fact indicates that, in principle, the moments can be  integrable with the distribution function $f_{15}$ (concerning the integrability of moments see \cite{Boillat-1997}). Nevertheless, for the non-linear  moment closure, there is the problematic that was noticed first by Junk \cite{Junk-1998} that the domain of definition of the flux in the last moment equation is not convex, the flux has a singularity, and the equilibrium state lies on the border of the domain of definition of the flux. To avoid this difficulties in the molecular extended thermodynamics approach,  we consider, as usual, the processes near equilibrium.  Then, we expand \eqref{fgen} around an equilibrium state in the following form:
\begin{align}\label{fgenE}
 \begin{split}
 &f_{15}=f_E\left(1-\frac{m}{k_B}\tilde{\chi}\right), \\
 &\tilde{\chi} = \tilde{\lambda}+C_i \tilde{\lambda}_i + C_iC_j\tilde{\lambda}_{ij} + \left(C^2 +\frac{2\mathcal{I}}{m}\right)\tilde{\mu} + \left(C^2+\frac{2\mathcal{I}}{m}\right)C_i\tilde{\mu}_i + \left(C^2+\frac{4\mathcal{I}}{m}\right)C^2 \tilde{\zeta},
 \end{split}
\end{align}
where a tilde on a quantity indicates its nonequilibrium part. In the following, for simplicity, we use the notation $f$ instead of  $f_{15}$. Although the expansion of the exponential \eqref{fgen}  is truncated at the first order with respect to the nonequilibrium variables, there exists the possibility to construct RET theories wit high expansion as was presented first by Brini and Ruggeri in \cite{BR}. The high order expansion has the advantage to have a larger domain of hyperbolicity \cite{BR2,BR3} and to reduce the magnitude of the sub-shock formation in the shock structure \cite{MR}.

Inserting \eqref{fgenE} into \eqref{conventional} and \eqref{defDelta}, we obtain the following algebraic relation for Lagrange multipliers:
\begin{align}
 & \tilde{\lambda} \rho + \frac{1}{3} \tilde{\lambda}_{ll}\hat{F}_{ll}^E +\tilde{\mu}\hat{G}_{ll}^E + \tilde{\zeta}\hat{H}_{llmm}^E = 0, \nonumber\\
 & \tilde{\lambda}_i \hat{F}_{ij}^E + \tilde{\mu}_i \hat{G}_{llij}^E=0, \nonumber \\
 & \tilde{\lambda}\hat{G}^E_{ll} + \frac{1}{3}\tilde{\lambda}_{ll}\hat{G}^E_{llmm} + \tilde{\mu}\hat{H}^E_{llmm} +\tilde{\zeta}\left(\hat{F}^E_{llmmnn} + 3\hat{J}^{1|E}_{llmm} +2\hat{J}^{2|E}_{ll} \right) =  0, \nonumber \\
  & \tilde{\lambda}\hat{F}^E_{ij} + \tilde{\lambda}_{rs}\hat{F}^E_{ijrs} + \tilde{\mu}\hat{G}^E_{llij} +\tilde{\zeta}\left(\hat{F}^E_{llmmij} + 2\hat{J}^{1|E}_{llij}\right)  = -\frac{k_B}{m}\left(\Pi \delta_{ij} - \sigma_{\langle ij\rangle}\right)  , \label{lamarg}\\
  & \tilde{\lambda}_i \hat{G}^E_{llij} + \tilde{\mu}_i\left(\hat{F}^E_{llmmij} + 2 \hat{J}^{E|1}_{llij} + \hat{J}^{E|2}_{ij}\right) =  - 2\frac{k_B}{m} q_j , \nonumber \\
  & \tilde{\lambda}\left(\hat{F}^E_{llmm} + 2\hat{J}^{E|1}_{ll}\right) + \frac{1}{3}\tilde{\lambda}_{ll}\left(\hat{F}^E_{llmmnn} + 2 \hat{J}^{E|1}_{llmm}\right) + \tilde{\mu}\left(\hat{F}^E_{llmmnn} + 3 \hat{J}^{E|1}_{llmm} + 2\hat{J}^{E|2}_{ll}\right) +\tilde{\zeta}\left(\hat{F}^E_{kkllmmnn} + 4\hat{J}^{E|1}_{llmmnn} + 4\hat{J}^{E|2}_{llmm}\right)  = -\frac{k_B}{m}\Delta, \nonumber
\end{align}
where $E$ with a quantity indicates the moment evaluated by the equilibrium distribution function, and 
\begin{align*}
   &\hat{J}^{1|E}_{k_1k_2 \cdots k_s}= \hat{G}^E_{llk_1k_2 \cdots k_s}- \hat{F}^E_{llk_1k_2 \cdots k_s} = \int_{\R^3} \int_0^{+\infty} 2 C_{k_1}C_{k_2} \cdots C_{k_s}  f_E \, \mathcal{I} \, \dA  ,\\
 &\hat{J}^{2|E}_{k_1k_2 \cdots k_t} = \int_{\R^3} \int_0^{+\infty} m  C_{k_1}C_{k_2} \cdots C_{k_t}f_E \left(\frac{2\mathcal{I}}{m}\right)^2 \dA .
\end{align*}

Taking into account the moments of $f^{I}_E $, i.e., \eqref{fint}, \eqref{int1e} and \eqref{cvI}, we obtain the following relation:
\begin{align*} 
	 &\hat{F}_{k_1k_2 \cdots k_r}^E = \hat{F}_{k_1k_2 \cdots k_r}^M,\\
  &\hat{J}^{1|E}_{k_1k_2 \cdots k_s}= 2 \frac{p}{\rho}{y}^I \hat{F}_{k_1k_2 \cdots k_s}^M,\\
 &\hat{J}^{2|E}_{k_1k_2 \cdots k_t} = 4 \frac{p}{\rho} \left(\hat{c}_v^I + {y^{I^2}} \right)\hat{F}_{k_1k_2 \cdots k_t}^M,
\end{align*}
where  $\hat{F}_{k_{1} k_{2} \ldots k_{r}}^{M}$ is the equilibrium moments for monatomic gas defined by
\begin{align}
 \hat{F}_{k_{1} k_{2} \ldots k_{r}}^{M} = m\int_{\mathbb{R}^{3}}f^K_{E}  C_{k_{1}} C_{k_{2}} \ldots C_{k_{r}}  d {\boldsymbol C}. \label{Fmom}
\end{align}
Since  $f^K_E$ is the Maxwell distribution \eqref{fKI}$_1$,  $\hat{F}_{k_{1} k_{2} \ldots k_{r}}^{M}$ are easily obtained, e.g.,
\begin{align*}
 &\hat{F}^M = \rho , \quad \hat{F}_{ij}^M = p\delta_{ij} , \quad \hat{F}_{ijrs}^M = \frac{p^2}{\rho}\left(\delta_{ij}\delta_{rs}+\delta_{ir}\delta_{js}+\delta_{is}\delta_{jr}\right),\\
 &\hat{F}_{llijrs}^M=7\frac{p^3}{\rho^2}\left(\delta_{ij}\delta_{rs}+\delta_{ir}\delta_{js}+\delta_{is}\delta_{jr}\right),\\
 &\hat{F}_{kkllmmnn}^M = 945 \frac{p^4}{\rho^3}.
\end{align*}

From \eqref{lamarg},  the intrinsic nonequilibrium Lagrange multipliers are evaluated as functions of $(\rho, T,\Pi,\sigma_{\langle ij\rangle},q_i,\Delta)$ up to the first order with respect to the nonequilibrium fields,  $\Pi$, $\sigma_{\langle ij\rangle}$, $q_i$ and $\Delta$. Instead of $\Delta$, it may be useful to introduce the following nonequilibrium field
\begin{align}
 \tilde{\Pi} = \frac{1}{3(4\hat{c}_v^I+5)} \left\{12\Pi (y^I+1) - \frac{\rho}{p}\Delta\right\}. \label{tilpi}
\end{align}
Then, we obtain as solution of \eqref{lamarg}:
\begin{align}
 \begin{split}
  & \tilde{\lambda}= \frac{3(\hat{c}_v^I -y^I)}{2\hat{c}_v^I \rho T}\Pi + \frac{3(4y^I+5)}{8\rho T}\tilde{\Pi},\\
  &\tilde{\lambda}_i = \frac{(2y^I+5)p}{p^2 T (2\hat{c}_v^I+5)} q_i,\\
  & \tilde{\lambda}_{ll} = - \frac{3(2\hat{c}_v^I +3)}{4\hat{c}_v^I p T} \Pi - \frac{3(y^I+1)}{2 pT}\tilde{\Pi},\\
  &\tilde{\lambda}_{\langle ij\rangle} = \frac{\sigma_{\langle ij\rangle}}{2pT}, \\
&\tilde{\mu}= \frac{3}{4\hat{c}_v^I p T}\Pi - \frac{3}{4 pT}\tilde{\Pi},\\
 &\tilde{\mu}_i = - \frac{\rho}{p^2 T (2\hat{c}_v^I+5)} q_i,\\
&\tilde{\zeta}=\frac{\rho}{8  p^2 T}\tilde{\Pi}.
 \end{split}
\label{Lagrange}
\end{align}
Inserting \eqref{mainE} and \eqref{Lagrange}
 into \eqref{lamdas},  we can write down the explicit form of the Lagrange multipliers.  As is well known, the multipliers coincide with the  main field
 \begin{align}
 \mathbf{u} ^\prime \equiv \left(\lambda,\lambda_i,\lambda_{ij},\mu,\mu_i,\zeta\right) \label{main}
\end{align}
 by which the system  \eqref{balanceFG1} becomes symmetric hyperbolic. Therefore we heave the well-posed Cauchy problem (local in time) ~\cite{RS,Boillat-1997}, and in some circumstances for small initial data, there exists global smooth solutions for all time (see ~\cite{book,newbook} and references therein).

 \subsection{Constitutive equations}

By using the distribution function \eqref{fgenE} with \eqref{Lagrange}, we obtain the constitutive equations for the fluxes up to the first order with respect to the nonequilibrium variables as follows:
\begin{align}
 \begin{split}
   &\hat{F}_{ijk}  
 = \frac{2}{2\hat{c}_v^I +5} (q_k \delta_{ij} + q_j \delta_{ik} + q_i \delta_{jk}),\\
 &\hat{G}_{llij}  
 = (2y^I+5)\frac{p^2}{\rho}\delta_{ij} + (2y^I + 7)\frac{p}{\rho}\Pi \delta_{ij} - (2\hat{c}_v^I +5)\frac{p}{\rho} \tilde{\Pi}  \delta_{ij} - (2y^I+7)\frac{p}{\rho} \sigma_{\langle ij\rangle},\\
 &\hat{H}_{llmmk} 
 = 20 \frac{p}{\rho} \frac{2y^I+ 2\hat{c}_v^I+7}{2\hat{c}_v^I + 5}q_k.  
 \end{split}
\label{CE}
\end{align}

\subsection{Nonequilibrium temperatures and generalized BGK model}

\subsubsection{Nonequilibrium temperatures}
We recall that $\varepsilon^K$ and $\varepsilon^I$ given in \eqref{conventional} are not equilibrium variables since these are the moments of the nonequilibrium distribution function (instead, the sum of the two is an equilibrium value). Then, we can define  two  nonequilibrium temperatures $(\theta^K,\theta^I)$ such that, by inserting in the equilibrium state function instead of $T$, we obtain the non equilibrium internal energies $(\varepsilon^K,\varepsilon^I)$, i.e.:   
\begin{align}
 \varepsilon^K = \varepsilon^K_E (\theta^K)=\frac{3}{2}\frac{k_B}{m}\theta^K, \qquad \varepsilon^I=\varepsilon_E^I(\theta^I). \label{noneqT}
\end{align}
Recalling  $2\rho \varepsilon^K = 3 \PP$ and \eqref{int1e} with \eqref{noneqT}$_1$, the total nonequilibrium pressure is expressed with $\theta^K$ from \eqref{EOS}$_1$ as follows:
\begin{align*}
 \PP = p(\rho,\theta^K) = \frac{k_B}{m}\rho \theta^K.
\end{align*}
Since $\PP = p+\Pi$, we have the following relations between the nonequilibrium temperature $\theta^K$ and the dynamical pressure $\Pi$:
\begin{align*}
 \theta^K = T\left(1+\frac{\Pi}{p(\rho,T)}\right).
\end{align*}
Moreover, we have the relation among three temperatures from \eqref{CaEqSt} and \eqref{calEOS} as follows:
\begin{align*}
 \varepsilon^I_E (\theta^I) - \varepsilon_E^I(T) = \varepsilon_E^K(T) - \varepsilon_E^K(\theta^K).
\end{align*}

\subsubsection{Generalized BGK model}

In polyatomic gases, we may introduce two characteristic times corresponding to two relaxation processes caused by the molecular collision:

(i) Relaxation time $\tau_K$: This characterizes the relaxation process within the translational mode (mode K) of molecules. The process shows the tendency to approach an equilibrium state of the mode K with the distribution function $f^K$ having the temperature $\theta^K$, explicit expression of which is shown below. However, the internal mode $I$ remains, in general, in nonequilibrium. This process exists also in monatomic gases.

(ii) Relaxation time $\tau$ of the second stage: After the relaxation process of the translational mode K, two modes, K and I, eventually approach a local equilibrium state characterized by $f_E$ with a common temperature $T$. Naturally we have assumed the condition: $\tau>\tau_K$.

To describe the above two separated relaxation processes, We adopt the generalized BGK collision term \cite{Struchtrup-1999,Struchtrup-2014} (see also \cite{Ruggeri-2020RdM,ET7,ET15}) which treats the translational relaxation and internal relaxation separately is as follows:
\begin{align}\label{bgkk}
 Q(f) = - \frac{1}{\tau_K}(f- f^K) - \frac{1}{\tau}(f-f_E),
\end{align}
where the distribution functions $f^K$ is
\begin{align*}
 f^K = \frac{\rho^I(\mathcal{I})}{m} \left(\frac{m}{2\pi k_B \theta^K}\right)^{3/2} \exp \left(-\frac{mC^2}{2k_B \theta^K}\right),
\end{align*}
with
\begin{align*}
 \rho^I(\mathcal{I}) = \int_{\R^3} mf d\cc.
\end{align*}

\subsubsection{Production terms}
From the generalized BGK model \eqref{bgkk}, the production terms given by \eqref{productioni} are evaluated as follows: 
\begin{align}
 \begin{split}
 &\hat{P}_{ll} = - \frac{3}{\tau}\Pi, \quad \hat{P}_{\langle ij\rangle} = \left(\frac{1}{\tau_K} + \frac{1}{\tau}\right)\sigma_{\langle ij\rangle}, \quad 
 \hat{Q}_{lli} = - 2 \left(\frac{1}{\tau_K} + \frac{1}{\tau}\right)q_i, \\
 &\hat{R}_{llmm} = -\left(\frac{1}{\tau_K}+\frac{1}{\tau}\right) \Delta + \frac{1}{\tau_K}\frac{p}{\rho} \Pi\left(12y^I +12 -3\frac{\Pi}{p}\right) .  
 \end{split}
 \label{prods}
\end{align}
Since we consider  linear constitutive equations, we neglect the quadratic term in the last expression of \eqref{prods}:
\begin{align*}
 \hat{R}_{llmm} &= -\left(\frac{1}{\tau_K}+\frac{1}{\tau}\right) \Delta + \frac{12}{\tau_K}\frac{p}{\rho}\left( y^I + 1\right) \Pi
  = - 12(y^I+1)\frac{p}{\rho} \frac{\Pi}{\tau} + 3(4\hat{c}_v^I+5)\frac{p}{\rho}\left(\frac{1}{\tau}+ \frac{1}{\tau_K}\right)\tilde{\Pi}.  
\end{align*}

\subsection{Closed field equations}

Using the constitutive equations above, we obtain the closed system of field equations for the 15 independent fields $(\rho, v_i, T, \Pi, \sigma_{\langle ij\rangle},q_i, \Delta)$ :
 \begingroup\makeatletter\def\f@size{8}\check@mathfonts
\def\maketag@@@#1{\hbox{\m@th\fontsize{10}{10}\selectfont\normalfont#1}}%
\begin{equation}
\begin{split}
&\frac{\partial \rho}{\partial t}+\frac{\partial}{\partial  x_i}( \rho v_i ) = 0, \\
 &\frac{\partial \rho v_j}{\partial t} + \frac{\partial }{\partial x_i}\left\{ [p+\Pi] \delta_{ij} -\sigma_{\langle ij\rangle}+ \rho v_i v_j\right\} = 0, \\
  &\frac{\partial}{\partial t} \left\{ p(2y^I +3) +\rho v^2 \right\} 
 +\frac{\partial}{\partial x_i} \left\{ 2q_i+  \left[p(2y^I + 5) +2\Pi\right]v_i - 2\sigma_{\langle li\rangle}v_l  + \rho v^2v_i\right\} =0,\\
 &\frac{\partial}{\partial t}\left\{ 3\left(p+\Pi\right) +\rho v^2 \right\} 
 + \frac{\partial}{\partial x_k}\left\{ \frac{10}{2\hat{c}_v^I+5}q_k + 5(p+\Pi)v_k -2\sigma_{\langle lk\rangle}v_l  +  \rho v^2v_k \right\} 
= -\frac{3\Pi}{\tau},\\
&\frac{\partial}{\partial t}\left(-\sigma_{\langle ij\rangle} + \rho v_{\langle i}v_{j\rangle}\right)
 +\frac{\partial}{\partial x_k}\left\{\frac{2}{1+\hat{c}_v} q_{\langle i}\delta_{j\rangle k} + 2[p+\Pi]v_{\langle i}\delta_{j\rangle k} - \sigma_{\langle ij\rangle}v_k - 2\sigma_{\langle k\langle i\rangle}v_{j\rangle} + \rho v_{\langle i}v_{j\rangle}v_k\right\} =  \left(\frac{1}{\tau_K}+\frac{1}{\tau}\right) \sigma_{\langle ij\rangle},\\
&\frac{\partial}{\partial t}\left\{2q_i +  \left[p(2y^I+5)+2\Pi\right]v_i -2 \sigma_{\langle li\rangle}v_l + \rho v^2 v_i\right\}+\\
 &\quad +  \frac{\partial}{\partial x_k} \Bigg\{ (2y^I+5)\frac{p^2}{\rho} \delta_{ik} + (2y^I + 7)\frac{p}{\rho}\Pi \delta_{ik} - \frac{p}{\rho} \frac{2\hat{c}_v^I +5}{3(4\hat{c}_v^I+5)} \left(12\Pi (y^I+1) - \frac{\rho}{p}\Delta\right)\delta_{ik}
 - (2y^I + 7)\frac{p}{\rho}  \sigma_{\langle ik\rangle} \\
 &\qquad \qquad \ + \frac{2}{2\hat{c}_v^I +5}q_lv_l \delta_{ik}+ 4 \frac{\hat{c}_v^I+3}{2\hat{c}_v^I+5}(q_iv_k + q_kv_i) 
 + (p+\Pi)v^2\delta_{ik} + \left[(2y^I+7) p + 4\Pi\right]v_iv_k \\
 &\qquad \qquad \ - \sigma_{\langle ik\rangle}v^2 - 2 \sigma_{\langle lk\rangle}v_lv_i - 2v_lv_k \sigma_{\langle il\rangle} + \rho v^2 v_i v_k\Bigg\} = -2 \frac{\Pi}{\tau}v_i  + 2 \left(\frac{1}{\tau_K}+\frac{1}{\tau}\right) {\sigma_{\langle il\rangle}} v_l  - 2 \left(\frac{1}{\tau_K}+\frac{1}{\tau}\right)q_i,\\
 &\frac{\partial}{\partial t}\left\{3\frac{p^2}{\rho}\left(4 y^I+5\right)
 +\Delta
 + 8v_i q_i + 2v^2 [p(2y^I+5) + 2 \Pi] - 4v_iv_j\sigma_{\langle ij\rangle}  + \rho v^4 \right\} + \\
 &\quad  + \frac{\partial}{\partial x_k} \Bigg\{ \frac{20p}{\rho(2\hat{c}_v^I+5)}\left(2\hat{c}_v^I + 2 y^I + 7\right) q_k
 +5 \left(4y^I+7\right)\frac{p^2}{\rho}v_k
 + 20 (y^I+2) \frac{p}{\rho}\Pi v_k-  \frac{5p(4\hat{c}_v^I + 7)}{3\rho \left(4c_v^I +5\right)}  \left[12\Pi (y^I+1) - \frac{\rho}{p}\Delta\right]v_k \\
 &\qquad \qquad \
  - 4 \left(2y^I + 7\right) \frac{p}{\rho} \sigma_{\langle ik\rangle}v_i
   + \frac{4(2\hat{c}_v^I + 7)}{2\hat{c}_v^I + 5}\left( q_k v^2 + 2 q_iv_i v_k\right) -4  \sigma_{\langle ik\rangle}v^2 v_i -4 \sigma_{\langle ij\rangle} v_i v_j v_k+ 2 \left[p(2y^I + 7)+4\Pi\right]v^2v_k + \rho v^4 v_k \Bigg\}\\
 &\qquad \qquad \    
 =  - 4v^2 \frac{\Pi}{\tau} + 4 \left(\frac{1}{\tau_K}+\frac{1}{\tau}\right) v_{\langle i}v_{j\rangle}{\sigma_{\langle ij\rangle}} + 8\left(\frac{1}{\tau_K}+\frac{1}{\tau}\right) v_i {q_i}
 - \left(\frac{1}{\tau_K}+\frac{1}{\tau}\right)\Delta + \frac{12}{\tau_K}\frac{p}{\rho} \left(y^I + 1\right)\Pi.
\end{split}
\label{balancekc}
\end{equation}
\endgroup
where, from \eqref{tilpi}, 
\begin{align}
\Delta   = 3\frac{p}{\rho} \left\{4 (y^I+1)\Pi - (4\hat{c}_v^I+5)\tilde{\Pi} \right\}. \label{tilpi-del}
\end{align}
In conclusion: 
\emph{The system \eqref{balancekc} formed by $15$ equations in the $15$ unknown  is closed with the provided equilibrium state function \eqref{EOS} and relaxation times $\tau$ and  $\tau_K$.
}

We remark that the field equations of $\rho, v_i, T, \Pi$ and $\sigma_{\langle ij\rangle}$ are same with the ones of polyatomic $14$ field theory and the presence of $\Delta$ involves only the last two equations of \eqref{balancekc} .

\subsection{Entropy density, flux and production}

The entropy density $h$ satisfies the entropy balance equation:
\begin{align*}
 \frac{\partial h}{\partial t} + \frac{\partial}{\partial x_i} (hv_i + \varphi_i) = \Sigma,
\end{align*}
where $\varphi_i$ is the non-convective entropy flux and $\Sigma$ is the entropy production which are defined below.

By adopting \eqref{fgenE} with \eqref{Lagrange}, we obtain the entropy density within second order with respect to the nonequilibrium variables 
\begin{align}\label{entropia}
 \begin{split}
 h 
 =&\rho s  - \frac{3(2\hat{c}_v^I+3)}{8\hat{c}_v^I p T} \Pi^2 - \frac{3(4\hat{c}_v^I + 5)}{16pT}\tilde{\Pi}^2
 -\frac{1}{4pT}\sigma_{\langle ij\rangle}\sigma_{\langle ij\rangle} - \frac{\rho}{(2\hat{c}_v^I+5)p^2 T} q_i q_i.
 \end{split}
\end{align}
This means that the entropy density is convex (in the limit of the approximation), it reaches the maximum at equilibrium and the system \eqref{balancekc} provides the symmetric form in the main field components.  
Similarly, the entropy flux is obtained as follows:
\begin{align*}
 \varphi_i  =& -k\int_{\R^3} \int_0^{+\infty} C_i f \log f\, \dA\\ 
 = & \frac{1}{T}q_i + \frac{2}{pT (2\hat{c}_v^I +5)}q_j \sigma_{\langle ij\rangle}  - \frac{2}{pT (2\hat{c}_v^I +5)}q_i \Pi + \frac{1}{pT}q_i \tilde{\Pi},
\end{align*}
The entropy production $\Sigma$ according with the symmetrization theorem \cite{RS,book,newbook} is obtained as scalar product between the main field given by \eqref{main} and the production vector given by \eqref{productioni}. By taking into account \eqref{prods} and \eqref{Lagrange},  we have
\begin{align*}
 \Sigma = & \mathbf{u} ^\prime \cdot \mathbf{f}  = \hat{\Sigma} =  \hat{\mathbf{u}} ^\prime \cdot \hat{\mathbf{f} } = \frac{\hat{\lambda}_{ll}}{3}\Pi  - \hat{\lambda}_{\langle ij\rangle}\sigma_{\langle ij\rangle} + 2 \hat{\mu}_i q_i + \hat{\zeta}\Delta\\
  =& \frac{3(2\hat{c}_v^I +3)}{4\hat{c}_v^I pT}\frac{1}{\tau}  {\Pi}^2 + \frac{3(4\hat{c}_v^I + 5)}{8 pT}\left(\frac{1}{\tau_K}+\frac{1}{\tau}\right) \tilde{\Pi}^2
 + \frac{1}{2pT}\left(\frac{1}{\tau_K}+\frac{1}{\tau}\right)\sigma_{\langle ij\rangle}\sigma_{\langle ij\rangle} + \frac{2\rho}{p^2T(2\hat{c}_v^I+5)} \left(\frac{1}{\tau_K}+\frac{1}{\tau}\right)q_i q_i.
\end{align*}
It is noteworthy that the entropy production is positive provided  the relaxation times are together with $\hat{c}_v^I\geq 0$.

\subsection{Characteristic velocities}
The differential system \eqref{balancekc} is particular case of a generic balance law system:
\begin{equation*}
\frac{\partial \, \mathbf{u}}{\partial t} + \frac{\partial\, \mathbf{F}^i(\mathbf{u})}{\partial x^i} =\mathbf{f}(\mathbf{u}),
\end{equation*}
and it is well known that the  characteristic velocity $V$ associated with a hyperbolic system of equations can be obtained by using the operator chain rule (see \cite{book}):
\begin{equation*}
\frac{\partial }{\partial t} \, \rightarrow - V \delta, \quad \frac{\partial }{\partial x_i} \, \rightarrow n_i\delta, \quad \mathbf{f}  \rightarrow 0,
\end{equation*} 
 where $n_i$ denotes the $i$-component of the unit normal to the wave front, $\mathbf{f}$ is the production terms and $\delta$ is a differential operator.  
 
 Let us consider only one dimensional space-variable, and the system \eqref{balancekc} reduces to only $7$ scalar equations for the $7$ unknown $(\rho,v=v_1,T,\Pi,\sigma=\sigma_{\langle 11\rangle}, q=q_1,\Delta)$.
 After some cumbersome calculations, it is possible to prove that the system  has  the following $7$ characteristic velocities evaluated in equilibrium:
\begin{align}\label{Uvdw}
\begin{split}
& V^{(1)} = v - U_E^{\text{1st}}\, \sqrt{\frac{k_B}{m}T}, \quad V^{(2)} =v-U_E^{\text{2nd}}\, \sqrt{\frac{k_B}{m}T}, \\ &  V^{(3)}= V^{(4)}=  V^{(5)} = v, \\ & V^{(6)} =v+U_E^{\text{2nd}}\, \sqrt{\frac{k_B}{m}T}, \quad V^{(7)} = v + U_E^{\text{1st}}\, \sqrt{\frac{k_B}{m}T},
\end{split}
\end{align}
with
\begin{align}
  &U_E^{\text{1st}} = \frac{1}{\sqrt{6(4 \hat{c}_v-1)(\hat{c}_v+1)}}\sqrt{ 8 \hat{c}_v (7 \hat{c}_v+11)-13+\sqrt{4 \hat{c}_v \left\{64 \hat{c}_v^2   (\hat{c}_v+16)+897\hat{c}_v-482\right\}+349}}, \label{U1max}\\
 &U_E^{\text{2nd}} = \frac{1}{\sqrt{6(4 \hat{c}_v-1)(\hat{c}_v+1)}}\sqrt{ 8 \hat{c}_v (7 \hat{c}_v+11)-13-\sqrt{4 \hat{c}_v \left\{64 \hat{c}_v^2   (\hat{c}_v+16)+897\hat{c}_v-482\right\}+349}}, \label{U2min}
\end{align}
where $\hat{c}_v = 3/2 + \hat{c}_v^I$ is the dimensionless specific heat. 
It is easy to prove that $U_E^{\text{1st}}$ given by \eqref{U1max} and $U_E^{\text{2nd}}$ by \eqref{U2min} are real because $\hat{c}_v\geqq 3/2$, and therefore the characteristic velocities \eqref{Uvdw} are all real in agreement that any symmetric systems are hyperbolic.

Note that the fastest  velocity $U_E^{\text{1st}} > U_E^{\text{2nd}}$ is larger than the corresponding one  of the polyatomic ET$_{14}$ theory, and this indicates that the subcharacteristic condition \cite{BoillatRuggeriARMA} is satisfied due to the convexity of entropy \eqref{entropia}. In the limit of monatomic gases ($\hat{c}_v = 3/2$), $U_E^{\text{1st}}\sim 2.27655$ and $U_E^{\text{2nd}} \simeq 1.16218$, which coincide with the ones of monatomic ET$_{14}$. In the limit that $\hat{c}_v \to \infty$, $U_E^{\text{1st}} \to\sqrt{3}$ which is same with the one predicted by polyatomic ET$_{14}$ in this limit. Recalling the general discussion of the dependence of the characteristic velocities on the degrees of freedom \cite{Arima-2014}, this is the value of the characteristic velocity of monatomic ET theory with $10$ moments (ET$_{10}$) in which $(F, F_i, F_{ij})$ are the only independent fields. On the other hand, in this limit, $U_E^{\text{2nd}} \to \sqrt {5/3}$ is different from the one of ET$_{14}$ but is same with the equilibrium sound velocity (the characteristic velocity of Euler system) of monatomic gases  in which $(F, F_i, F_{ll})$ are the only independent fields.  While, $U_E^{\text{2nd}}$ of ET$_{14}$ approaches to $1$ which is the characteristic velocity of ET theory with $4$ moments (ET$_{4}$) in which $(F,F_i)$ are the only independent fields.

\begin{figure}[h!]
	\centering
	\includegraphics[width=77mm]{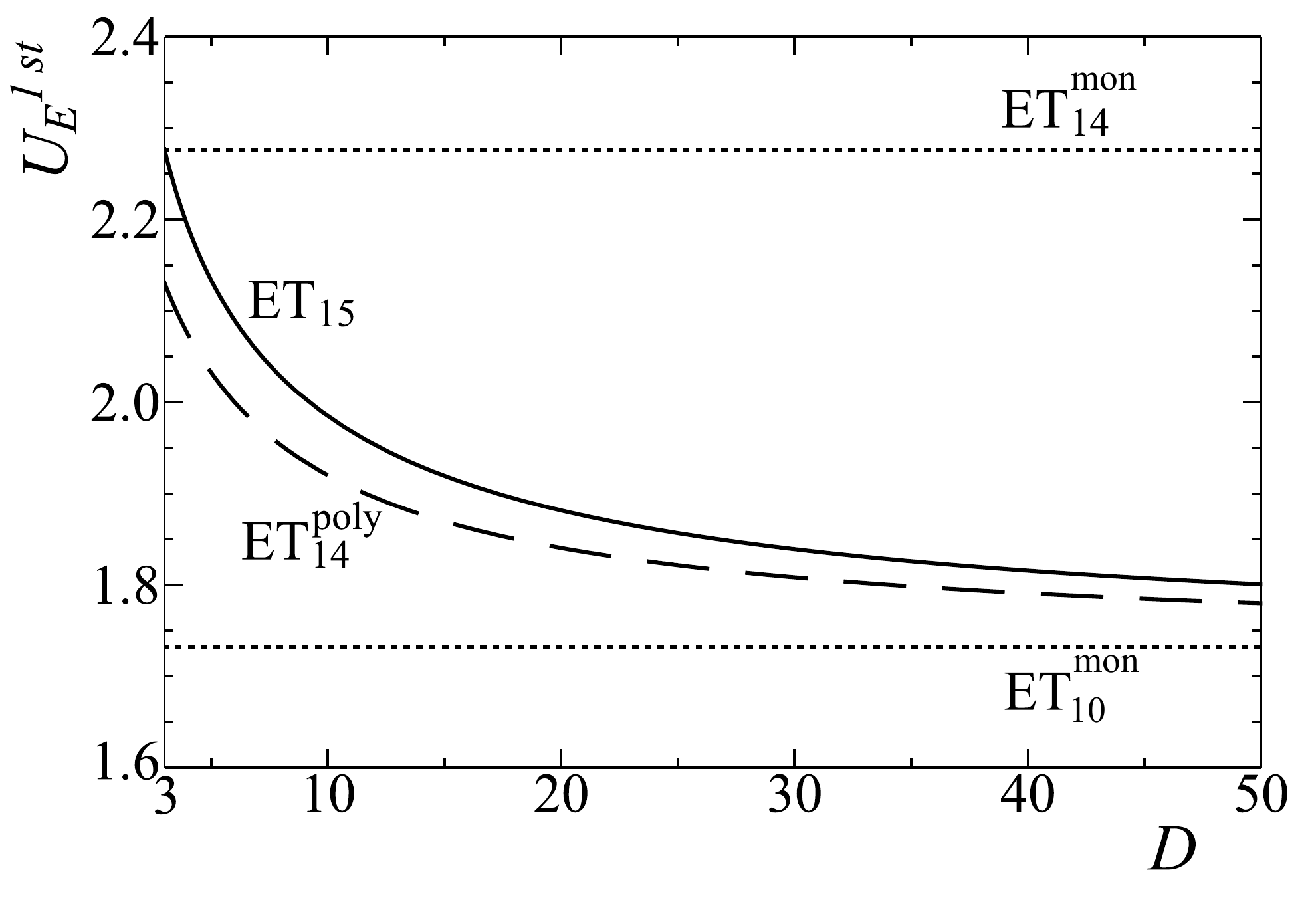}
 	\includegraphics[width=77mm]{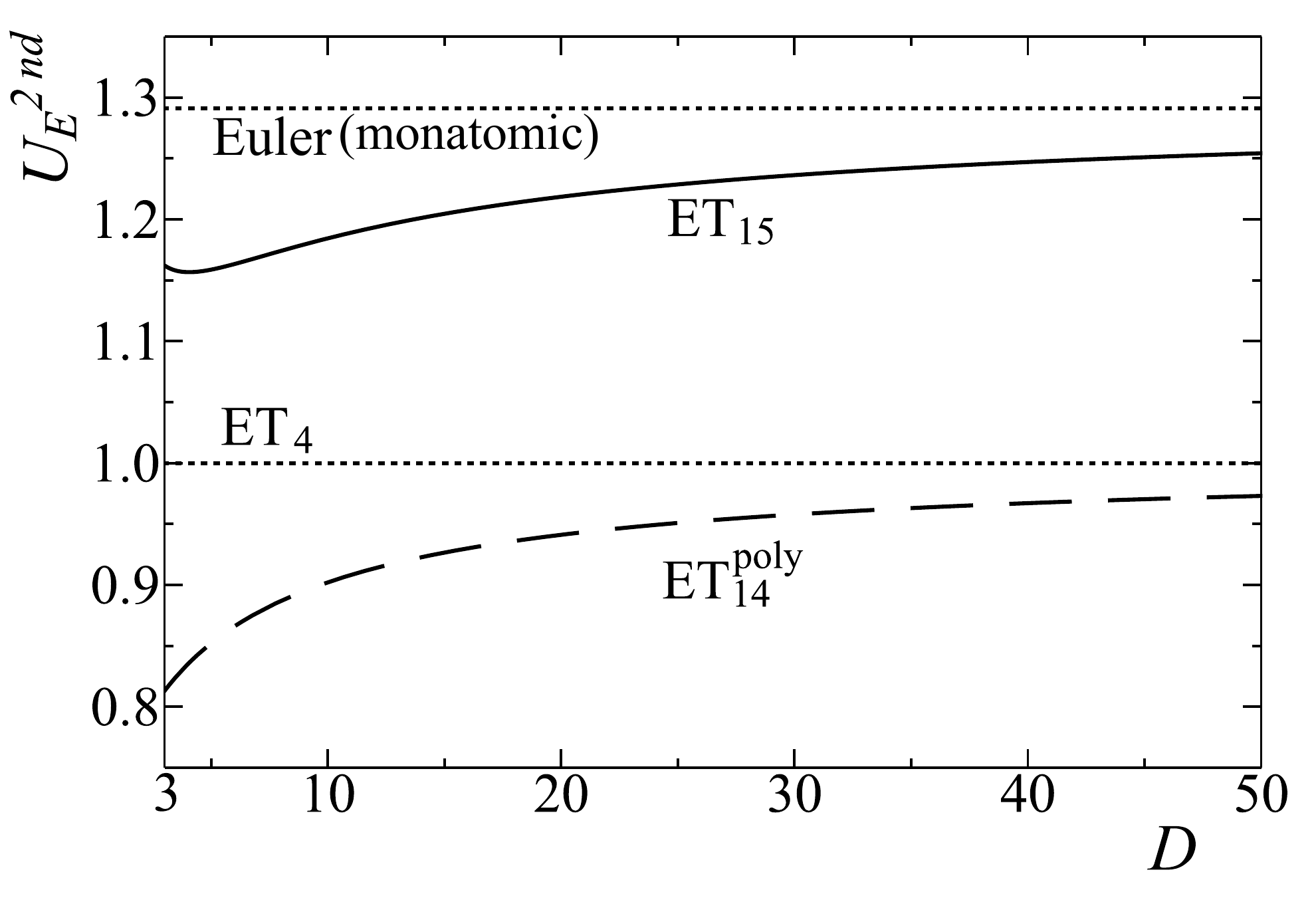}
	\caption{Dependence of the normalized characteristic velocities $U_E^{\text{1st}} $ (left) and $U_E^{\text{2nd}}$ (right) with respect to  $D$. The solid and dashed lines are the normalized characteristic velocities of ET$_{15}$ and polyatomic ET$_{14}$. The limit value of the normalized characteristic velocities of $D\to 3$ and $D\to \infty$ are indicated with dotted lines.  In the limit that $D\to 3$,  $U_E^{\text{1st}} $ and $U_E^{\text{2nd}}$ of ET$_{15}$ coincide with the monatomic ET$_{14}$. In the limit that $D\to \infty$, both of $U_E^{\text{1st}} $ of ET$_{15}$ and ET$_{14}$ approach to the one of monatomic ET$_{10}$, and $U_E^{\text{2nd}} $ of ET$_{15}$ and ET$_{14}$ approach, respectively, to monatomic Euler and ET$_{4}$.}
	\label{Fig:CV}
\end{figure}

In the case of the polytropic gas of which equations of state are given in \eqref{EOSnonpoly}, $\hat{c}_v =D/2$,  the normalized characteristic velocities  $U_E^{\text{1st}}, U_E^{\text{2nd}}$  depend only on the degrees of freedom $D$. The dependences are shown in Fig.\ref{Fig:CV}.

\subsection{Maxwellian iteration and phenomenological coefficients}

The NSF theory is obtained by carrying out the Maxwellian iteration \cite{Ikenberry} on \eqref{balancekc} in which only the first order terms with respect to the relaxation times are retained. Then we obtain
\begin{align}
 &\Pi = - p\tau  \frac{4\hat{c}_v^I}{6\hat{c}_v^I + 9} \frac{\partial v_k}{\partial x_k}, \quad
 \sigma_{\langle ij\rangle} = 2p \tau_\sigma \frac{\partial v_{\langle i}}{\partial x_{j \rangle}}, \quad
 q_i = - p\tau_q  \frac{2\hat{c}_v^I+5}{2} \frac{k_B}{m} \frac{\partial T}{\partial x_i}, \label{NFS1} 
\end{align}
and
\begin{align}
 \Delta = -\tau_\Delta \frac{16\hat{c}_v^I}{2\hat{c}_v^I + 3}\frac{p^2}{\rho} \left(y^I+1\right)\left(1+\frac{\tau}{\tau_K }\right)\frac{\partial v_l}{\partial x_l}, \label{Del_max}
\end{align}
where
\begin{align*}
 \frac{1}{\tau_\sigma } =  \frac{1}{\tau_q} =  \frac{1}{\tau_\Delta}= \frac{1}{\tau_K}+ \frac{1}{\tau}.
\end{align*}
Recalling the definition of the bulk viscosity $\nu$, shear viscosity $\mu$, and heat conductivity $\kappa$ in the NFS theory:
\begin{align}\label{NFSs}
 \Pi =-\nu \frac{\partial v_i}{\partial x_i},
 \qquad
 \sigma_{\langle ij\rangle} = 2 \mu \frac{\partial v_{\langle i}}{\partial x_{j\rangle}}, \qquad
 q_i = - \kappa \frac{\partial T}{\partial x_i},
\end{align}
we have from \eqref{NFS1}
\begin{align}
 \begin{split}
 &\nu =  \frac{4\hat{c}_v^I}{6\hat{c}_v^I + 9} p \tau  ,\qquad
 \mu = p \tau_\sigma,   \qquad 
 \kappa = \frac{2\hat{c}_v^I+5}{2} p \tau_q.  
 \end{split}
\label{maxwellite}
\end{align}
We note that $\Delta$ and $\tilde{\Pi}$ is not present in the conservation laws of mass, momentum and energy. In particular, \eqref{Del_max} indicates with \eqref{tilpi-del} 
\begin{align*}
 \tilde{\Pi}=0.
\end{align*}
This result seems similar to the case of monatomic ET$_{14}$ in which the nonequilibrium scalar field is equal to $0$ in the Maxwellian iteration \cite{Kremer14}.


As usual in the BGK model, the Prandtl number predicted by the present model is not satisfactory. To avoid this difficulty, one possibility is to regard the relaxation times $\tau$, $\tau_\sigma$ and $\tau_q$ as functions of $\rho$ and $T$, and estimate them by using the experimental data on $\nu$, $\mu$ and $\kappa$. On the other hand, $\tau_\Delta$ and $\tau_K$ are not related to such phenomenological coefficients and the kinetic theory is needed for their estimation,  or we may determine these relaxation times  as parameters  to have a  better agreement with some experimental data as it has been usually done for the bulk viscosity.
\medskip

Summarizing we have the following result:
\emph{With the Maxwellian iteration procedure,  the hyperbolic  system \eqref{balancekc} converges (in similar way of the $14$ fields theory) to the classical parabolic system of NFS formed by the first five  equations of \eqref{balancekc} with  the constitutive equations \eqref{NFSs} with bulk and shear viscosities and heat conductivity related to the relaxation times by \eqref{maxwellite}.}

\subsection{Principal subsystem}

The concept of the principal subsystem for a general system of hyperbolic system of balance laws was  introduced in \cite{BoillatRuggeriARMA}.  By definition, some  components of the  main field are put as a constant and the corresponding balance laws are deleted. In this way, we have a small set of the field equations from a large set of the field equations that has the property that the  entropy principle is preserved and the sub-characteristic conditions are satisfied, i.e., the spectrum of characteristic eigenvalues of the small system is contained in the spectrum of the larger one. As consequence, in the moments theory, the maximum characteristic velocity increases with the number of moments \cite{Boillat-1997}. 

In the present case, the polyatomic ET$_{14}$ is obtained as a principal subsystem of ET$_{15}$ under the condition $\zeta=0$, i.e., from \eqref{Lagrange}$_7$,
\begin{align*}
 \tilde{\Pi} = 0,
\end{align*}
or, in other words,
\begin{align*}
 \Delta = 12\frac{p}{\rho}\Pi \left(y^I+1\right),
\end{align*}
and \eqref{balancekc}$_7$ is ignored.

\subsection{Monatomic gas limit}

The monatomic gases are described in the limit $\varepsilon^I \to 0$ ($y^I \to 0$) and therefore $\hat{c}_v^I \to 0$.
In the limit, the equation for $\Pi$ obtained by subtracting \eqref{balancekc}$_3$ from \eqref{balancekc}$_4$ becomes
\begin{align}
   \label{14pi}
&\frac{\partial \Pi}{\partial t} + v_k\frac{\partial \Pi }{\partial x_k}
    = - \left(\frac{1}{\tau_\Pi} + \frac{\partial v_k}{\partial x_k}\right) \Pi.
\end{align}
This is the first-order quasi-linear partial differential equation with respect to $\Pi$. 
As it has been studied in \cite{Arima-2013}, the initial condition for \eqref{14pi} must be compatible with the case of monatomic gas, i.e., $\Pi(0, \boldsymbol{x}) = 0$, and, assuming the uniqueness of the solution, the possible solution of Eq. \eqref{14pi} is given by
\begin{align}
   \label{Pi0}
   &\Pi(t, \boldsymbol{x}) = 0 \ \ \ (\textrm{for any} \ t).
\end{align}

If we insert the solution \eqref{Pi0} into \eqref{energies} and \eqref{CE} with $y^I=0$ and $\hat{c}_v^I=0$, the velocity independent moments are expressed by the velocity independent moments of monatomic gas $\hat{F}_{i_1i_2 \cdots i_n}^M$  which are given in \eqref{Fmom} as follows:
\begin{align*}
	&\hat{F}_{ij} = p\delta_{ij} - \sigma_{\langle ij\rangle} = \hat{F}_{ij}^M,\\
 &\hat{F}_{ijk} = \frac{2}{5}\left(q_i \delta_{jk} + q_j\delta_{ik} + q_k \delta_{ij}\right) = \hat{F}_{ijk}^M,\\
 &\hat{G}_{ll} = \hat{F}_{ll} = 2 \rho \varepsilon^K = 3p = \hat{F}_{ll}^M,\\
  &\hat{G}_{lli}=\hat{F}_{lli} = 2q_i = \hat{F}_{lli}^M,\\
 &\hat{G}_{llij} = \frac{5p^2}{\rho}\delta_{ij} + \frac{1}{3}\Delta \delta_{ij} - 7 \frac{p}{\rho}\sigma_{\langle ij\rangle} = \hat{F}_{llij}^M,\\
 &\hat{H}_{llmm} = 15\frac{p^2}{\rho} + \Delta = \hat{G}_{llmm}=\hat{F}_{llmm}^M,\\
 &\hat{H}_{llmmk} = 28 \frac{p}{\rho} q_k=\hat{F}_{llmmk}^M.
\end{align*}
Then, the $F$'s hierarchy coincides with the monatomic $F$'s hierarchy and $G$'s and $H$'s hierarchies coincide with the corresponding  monatomic $F$'s hierarchy.
This indicates that, in this singular limit,  solutions of ET$_{15}$ converge to those of  monatomic $14$ theory by Kremer \cite{Kremer14}.

It may be remarkable that, from this coincidence, we can set the inessential phenomenological constants appear in monatomic  ET$_{14}$ theory \cite{Kremer14} as zero. Therefore, the molecular approach can determine the constitutive equations without arbitrariness except for the production terms, although the phenomenological approach can provide the theory for the gas with generic equations of state, e.g., the theory for degenerate  Fermi and Bose gases \cite{Kremer14}.

\section{Dispersion Relation}

The dependences of the phase velocity and the attenuation per wavelength on the frequency are studied. 

\subsection{Phase velocity and attenuation factor}

Let us confine our study within one-dimensional problem, that is, a plane longitudinal wave propagating in $x$-direction.  Therefore, considering the symmetry of the wave, we have the following form: 
\begin{align}
 &v_i \equiv \left(
 \begin{array}{c}
  v\\ 0\\ 0
   \end{array}
 \right), \ \ \ 
 \sigma_{\langle ij\rangle} \equiv \left(
 \begin{array}{ccc}
  \sigma & 0 &0\\
   0& -\frac{1}{2}\sigma & 0\\
  0& 0 & -\frac{1}{2}\sigma
   \end{array}
 \right),\ \ \ 
 q_i \equiv \left(
 \begin{array}{c}
  q\\ 0\\ 0
   \end{array}
 \right). 
 \label{1dim}
\end{align}
Moreover, we study a harmonic wave for the fields $\boldsymbol{u}=(\rho, v, T, \Pi, \sigma, q, \Delta)$ with the angular frequency $\omega$ and the complex wave number $k$ such that 
\begin{align}
 & \boldsymbol{u}=\boldsymbol{w}  \mathrm{e}^{\mathrm{i} (\omega t - k x)},   \label{harmonic}
\end{align}
where $\boldsymbol{w}$ is a constant amplitude vector.
From Eq. \eqref{balanceFG1} with \eqref{1dim} and \eqref{harmonic}, the dispersion relation $k=k(\omega)$ is obtained by the standard way \cite{RET}.  
The phase velocity $v_{ph}$ and the attenuation factor $\alpha$ are calculated as the functions of the frequency $\omega$ by using the following relations:
\begin{align*}
 v_{ph}= \frac{\omega}{{\cal R}e(k)}, \quad \alpha = -  {\cal I}m(k).
\end{align*}
In addition, it is useful to introduce the attenuation per wavelength $\alpha_{\lambda}$:
\begin{align*}
 & \alpha_{\lambda}(\omega)= \alpha \lambda=\frac{2\pi v_{ph} \alpha }{\omega} = -2\pi \frac{{\cal I}m (k)}{{\cal R}e (k)},
\end{align*}
where $\lambda$ is the wavelength.

Let us introduce the following dimensionless parameters:
\begin{align*}
 &\Omega = \tau \omega ,  \ \ \ \hat{\tau}_K =  \frac{\tau_K}{\tau}.
\end{align*}
Then the dispersion relation depends on these parameters with dimensionless specific heat of internal mode $\hat{c}_v^I$. We emphasize that $k=k(\omega)$ does not depend on $\rho$, and its temperature dependence is determined through the dimensionless specific heat that can be determined from statistical mechanics or experimental data.  

\begin{figure}[h!]
	\centering
	\includegraphics[width=77mm]{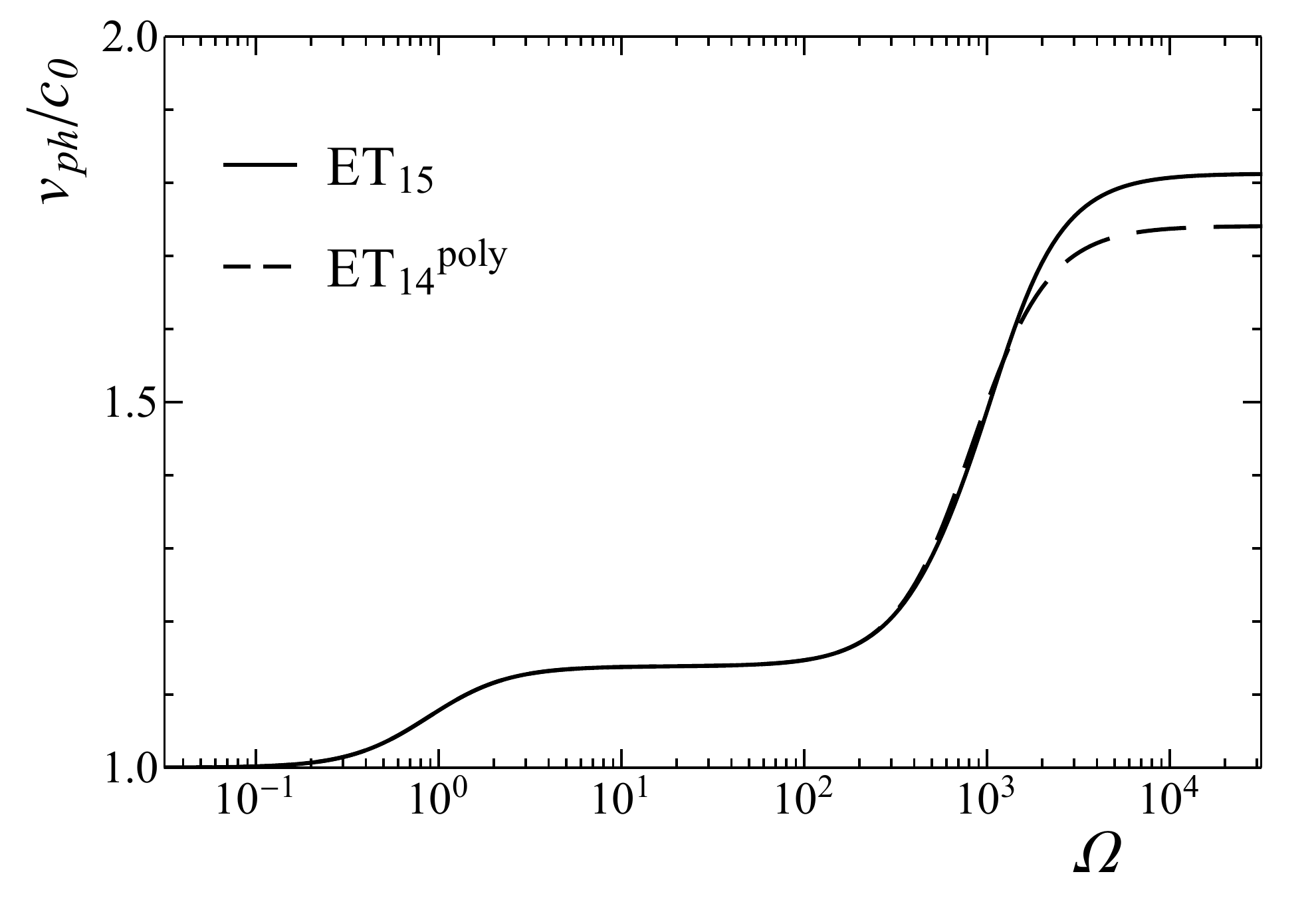}
	\includegraphics[width=77mm]{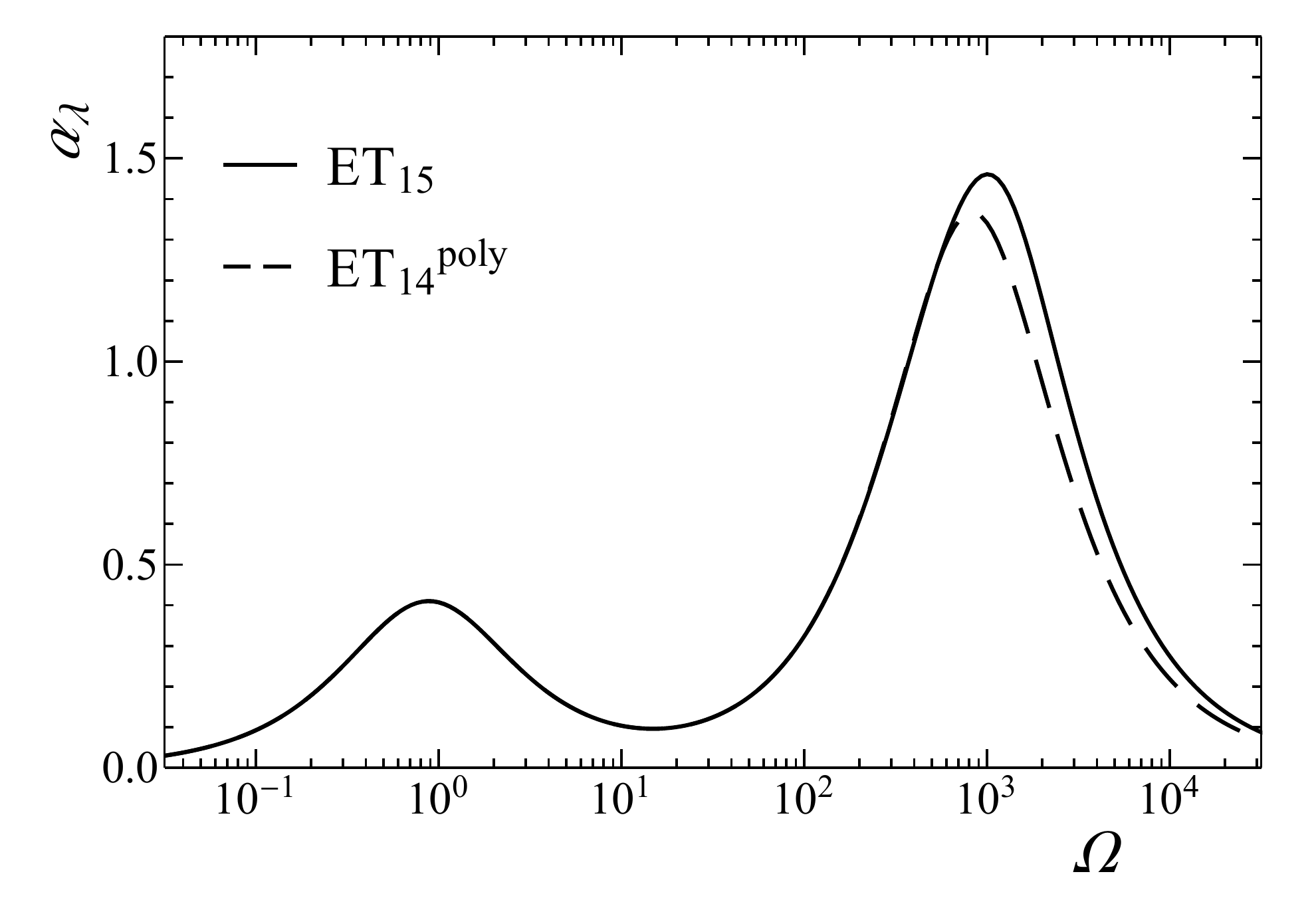}
	\caption{Typical dependence of the dimensionless phase velocity $v_{ph}/c_0$ (left) and the attenuation per wavelength $\alpha_\lambda$ (right) on the dimensionless frequency $\Omega$ predicted by the ET$_{15}$ and polyatomic ET$_{14}$ theories. }
	\label{Fig:dispersion}
\end{figure}

As an example, we adopt $\hat{\tau}_K = 0.001$ which indicates the existence of the slow relaxation of internal mode \cite{ET7,ET14linear,Arima-2019}. We show the dependence of the phase velocity normalized by the equilibrium sound velocity $c_0$:
\begin{align*}
 c_0 = \sqrt{\frac{2\hat{c}_v^I + 5}{2\hat{c}_v^I +3}\frac{k_B}{m}T},
\end{align*}
and the attenuation per wavelength on the dimensionless frequency in Fig. \ref{Fig:dispersion} in the case with $\hat{c}_v^I=2$. Around $\Omega \sim 10^0 (\omega \sim \tau^{-1})$, we can observe a change of $v_{ph}$ and a peak of $\alpha_\lambda$. Since this is due to the relaxation of internal mode relating to $\Pi$, both of the predictions by ET$_{14}$ and ET$_{15}$ coincide each other. Around $\Omega \sim 10^3 (=\hat{\tau}_K^{-1})$, we can observe a steep change of $v_{ph}$ and a large peak of $\alpha_\lambda$. Since this is due to the relaxation of $\sigma, q$ and $\Delta$, the difference between two theories emerges.

\section{Conclusions}
According to the general results of Pennisi and Ruggeri \cite{PRS}, the classical limit of the relativistic theory of moments provides more complex hierarchy than the $F$' and $G$' binary hierarchy. In this paper, we have studied the case of $N=2$ in which the classic limit dictates $15$ fields for a non-polytropic polyatomic gas. We have obtained the closure using the MEP. The closed field equations include the classical NSF theory as its parabolic limit and converge to the monatomic ET$_{14}$ theory obtained by Kremer \cite{Kremer14} in the monatomic singular limit.  Moreover, we proved that the polyatomic ET$_{14}$ theory is a principal subsystem of the present one, and according to the general results, the spectrum of characteristic velocities of ET$_{15}$ includes the spectrum of eigenvalues of ET$_{14}$. Finally, we have evaluated the dispersion relation proving that, in the low-frequency region, the predictions by ET$_{14}$ and ET$_{15}$ theories coincide with each other, while, in the high-frequency region, the difference between two theories emerges due to the existence of the additional higher order moment.

We finally remark that, in the present approach, we treat the internal modes as a whole; however, in principle, we can generalize the theory with two internal modes, one for the rotational and one for the vibrational motion of a molecule, as was done in the paper \cite{ET7,ET15}.

%



\section*{Acknowledgments} \label{sec:acknowledgements}
This paper is dedicated to the memory of Carlo Cercignani. The  work has been partially  supported by GNFM/INdAM (TR).

\vspace{6pt} 

\authorcontributions{All authors were fully involved in substantial conception and design of the paper; drafting the article and revising it critically for important intellectual content; final approval of the version to be published.
}


\funding{This work was partially supported by JSPS KAKENHI Grant Numbers JP18K13471 (Takashi Arima).}


\conflictsofinterest{The authors declare no conflict of interest.}

\reftitle{References}

\end{document}